\documentclass{article}[12]

\def\ref{\par\hangindent=20pt\hangafter=1\noindent}
\usepackage{graphicx}
\usepackage{pgf}
\usepackage{pgfpages}
\usepackage{authblk} 
\usepackage[longnamesfirst,authoryear]{natbib} 
\bibpunct{(}{)}{;}{a}{}{;}
\usepackage{enumitem}
\usepackage{lineno}
\usepackage{caption}
\usepackage{subcaption}

\newcommand{\bea}{\begin{eqnarray}}
\newcommand{\eea}{\end{eqnarray}}
\newcommand{\be}{\begin{equation}}
\newcommand{\ee}{\end{equation}}

\title{A Commentary on Statistical Assessment \\ of Violence Recidivism Risk}
\author[$^1$]{Peter B. Imrey}  
\author[$^2$]{A. Philip Dawid} 
\affil[$^1$]{Cleveland Clinic Lerner College of Medicine of Case Western Reserve University Cleveland, Ohio 44195, USA. \tt imreyp@ccf.org} 
\affil[$^2$]{Statistical Laboratory, University of Cambridge, Cambridge CB3 0WB, UK. \tt apd@statslab.cam.ac.uk}
\date{March 12, 2014}

\begin{document}
\maketitle

\begin{abstract}
  Increasing integration and availability of data on large groups of
  persons has been accompanied by proliferation of statistical and
  other algorithmic prediction tools in banking, insurance, marketing,
  medicine, and other fields (see e.g.,
  \cite{Steyerberg,SteyerbergEoMDM}).  Controversy may ensue when such
  tools are introduced to fields traditionally reliant on individual
  clinical evaluations.  Such controversy has arisen about
  ``actuarial'' assessments of violence recidivism risk, i.e., the
  probability that someone found to have committed a violent act will
  commit another during a specified period.  Recently
  \cite{HartMichieCooke} and subsequent papers from these authors in
  several reputable journals have claimed to demonstrate that
  statistical assessments of such risks are inherently too imprecise
  to be useful, using arguments that would seem to apply to
  statistical risk prediction quite broadly.  This commentary examines
  these arguments from a technical statistical perspective, and finds
  them seriously mistaken in many particulars. They should play no
  role in reasoned discussions of violence recidivism risk assessment.
\end{abstract}

\section{Introduction}
The prospect of violence is a substantial consideration for legal
decisions on bail, sentencing, parole, preventive confinement, and
liability of mental health professionals.  Courts receive information
on a person's propensity to commit violence from expert
testimony. Experts vary in how they approach this task, with some
emphasizing subjective clinical judgment and others preferring more
standardized methods. Structured risk assessments that integrate
selected sociodemographic, personal history, and psychometric
characteristics of the individual, and whose results have been
statistically associated with future violence in follow-up studies of
groups, have thus come into common use.  Among these, ``actuarial risk
assessment instruments'' (ARAIs) produce numerical estimates of a
probability of violent behavior that are analogous, in their
development and interpretation, to predicted probabilities by which
insurers ``rate'' clients and price policies for losses from
automotive accident, theft, extreme weather, disease, or death
\citep{YangWongCoid}.

Controversies about the respective values of clinical expertise and
explicit decision rules are chronic in many fields, including forensic
risk assessment. However, recent discussions of violence recidivism
(\citealt{HartMichieCooke}, henceforth HMC; \citealt{CookeMichie} with
erratum \citeyear{CookeMichieCorrig}, \citealt{CookeMichieVRA},
\citeyear{CookeMichieWhatWhy}, henceforth respectively CM1, CM2, CM3,
and CM1-3 collectively), and \citealt{HartCooke}, henceforth HC; are
exceptional in challenging actuarial risk prediction on technical
statistical grounds, turning the usual discourse on its head.  CoHaMi
(employed to jointly reference these when addressing their common
threads) use illustrative data and simulations derived from five ARAIs
(the Violence Risk Appraisal Guide (VRAG, \citealt{VRAG}), Psychopathy
Checklist Revised (PCL-R, \citealt{PCL-R}), Static-99
\citep{STATIC-99}, the Risk Matrix 2000 \citep{RM2000}, and a new ARAI
based on the Sexual Violence Risk-20 instrument (SVR-20,
\citealt{SVR-20})), to contest the utility of ARAI-based risk
predictions of violence recidivism generally. Their critique,
following closely on the heels of a caution by prominent statisticians
against over interpretation of survival time predictions for
individual medical patients \citep{HendersonKeiding}, contends that i)
ARAI-based risk assessments are unreliable due to misclassification
errors in risk category assignments (CM1), and ii) statistical
confidence intervals and prediction intervals for ``group and
individual risk estimates'' are inevitably, \emph{inherently}, too
wide for such instruments to be useful (HMC, CM1-3, HC). Objections
have been raised to these claims (\citealt{HarrisRiceQuinsey},
\citealt{MossmanSellke}, \citealt{HansonHoward},
\citealt{SkeemMonahan}, \citealt{ScurichJohn}), but these objections
have for the most part been rejected (\citealt{HartMichieCookeReply1,
  HartMichieCookeReply2}, CM1-3, HC), with the continuing and
frequently cited exchange providing fuel for further criticism of
statistical risk assessment in the forensic psychology and legal
communities (e.g., \citealt{Coyle, Starr}).

While the gravity and potential societal impact of legal decisions
influenced by recidivism risk prediction are high, the technical
arguments raised in this context might also be applied, at least in
principle, to predictions that inform medical prognosis and clinical
decisions, geological explorations for oil and natural gas, insurance
rating, marketing, loan rating, and athletic management and coaching
strategies.  If the critique is valid, its ramifications thus extend
into many areas of biomedical, public health, and behavioral science
investigation, and to standards of practice in several professions,
including applied statistics. This adds urgency to the need for
clarification.

Our purpose here is to identify crucial technical errors in HMC's,
CM1-3's, and HC's technical statistical arguments. Some aspects and
consequences of these errors have been raised by or are implicit in
previous commentaries, and one point has been partially conceded in
\cite{HartMichieCookeReply1} and sequelae. Moreover, the controversy
fueled by HMC and CM1-3 has largely stimulated the recent
\emph{Special Issue: Methodological Issues in Measuring and
  Interpreting the Predictive Validity of Violence Risk Assessments}
of the journal \emph{Behavioral Sciences and the Law}, in which HC is
accompanied by much instructive and useful discussion of both
technical and philosophical aspects of risk assessment. Nevertheless,
while this special issue and accumulating citations of these articles
reflect the seriousness with which HMC's and CM1-3's claims have been
received by some forensic psychologists, psychiatrists, and jurists,
the core statistical misconceptions and misapplications of
conventional formulae underlying their claims have yet to be directly
addressed.  Explicit refutation is desirable to forestall such
mistakes from seeding further unproductive argumentation in this and
other areas where similar statistical tools have been found useful.

The current commentary is thus most specifically pertinent to forensic
psychologists, psychiatrists, lawyers, and judges, for whom violence
recidivism risk assessments contribute to clinical recommendations and
legal decisions. The conceptual discussion of the nature of risk, and
statistical inferences about risk, should also be useful to others
concerned with public policy and/or professional practice in
circumstances requiring case-by-case judgments of prospects for other
human behaviors, illnesses, or misadventures. Since statistical
training in case-oriented disciplines may be minimal, the exposition
assumes no statistical background. However, comprehending the
fallacies involved requires understanding some basic statistical
technical concepts, which are presented with minimal mathematical
notation. Statistically knowledgeable readers, who may wish only to
skim such sections, should find this a disturbing case study, and may
benefit from the painstaking conceptual and terminological delineation
of alternative meanings of ``risk.'' These seem conflated in usage no
less often by statisticians as by others at the price, as in the
violence recidivism debate, of much confusion. For a thorough
discussion of the variety of conceptions of the term Òindividual
riskÓ, see \citealt{DawidIR}.

This paper takes no position on the proper role of ARAIs in recidivism
risk assessment. The hope is simply to clear specious statistical
arguments from the discourse, so discussions of ARAIs and recidivism
risk prediction are more usefully directed. The exposition throughout
refers to specific contexts or examples, usually from criminal cases,
to illustrate and make abstract points more concrete. No arguments
herein are specific to the civil or criminal context in which ARAIs
might be used, or to their use in predicting risk of recidivism as
distinguished from risk of initial violence, provided that use of any
particular ARAI conforms to the basic conditions of its development.

We focus on the specific meanings and uses of terms, formulae, and
computer simulations by HMC, CM1-3, and HC. Consistent with this
focus, remarks are confined to frequentist statistical inference,
simply because the methods applied controversially by HMC, CM1-3, and
HC all derive from frequentist assumptions, as do the properties that
have predominantly been used to justify more established uses of these
methods. It is thus most straightforward to address the problematic
issues on the specific turf where they arise. For purposes more
general than this paper, a Bayesian perspective has much to offer;
see, e.g., \citealt{DonaldsonWollert}, \citealt{ScurichJohn}, and
\citealt{HarrisRice},

Section 2.1 describes HMC's Table 1 of recidivism proportions and
associated intervals for nine risk strata, and the conclusions HMC
draw from them. The technical bases of frequentist probabilistic risk
prediction (Section 2.2) and statistical intervals (Section 2.3) are
then described and related to such data in a tutorial,
non-mathematical style. It is necessary to define, albeit informally,
general terms such as ``population'' and ``sample,'' ``parameter'' and
``statistic,'' ``risk'' and ``individual risk,'' ``estimation'' and
``prediction,'' rather than leaving such definitions implicit. While
this may seem pedantic to statistician readers, such specificity is
needed to cut through the semantic confusion. Section 3 describes the
technical fallacies on which CoHaMi's statistical objections to
ARAI-based risk assessment rely.  Section 4 reflects on the underlying
misperception that has fueled this controversy, identifies more
legitimate grounds for questioning the legal use of ARAIs, and points
towards more appropriate empirical approaches for evaluating their
performance.

\section{Defining Terms}
\subsection{Data and context}
Table 1, derived from Table 1 of HMC, exhibits the type of data from
which, in principle, ARAI-based risk prediction might proceed.  The
Violence Risk Appraisal Guide (VRAG) was administered to a total of
608 violent offenders, of whom 192 (31.6\%) committed a further
violent act in an ensuing unconfined period of up to ten years. The
table subdivides these offenders into categories of increasing VRAG
score, which is both theoretically and empirically correlated -- the
latter easily seen by scanning down the test score categories in Table
1 -- with an increasing proportion of five-year recidivists.

\begin{center} TABLE 1 HERE \end{center}

The general idea is then to use a new offender's VRAG score to predict
his or her risk of violence if left unconfined over the next five
years. The numerical prediction itself might be the proportion of
recidivists within the same test score category (row) of Table 1, or
an alternative ``smoothed'' value derived from a statistical model,
and which thus incorporates information from offenders in other VRAG
categories, filtered through assumptions of that model.  Proponents of
ARAI-based risk prediction argue that such numerical values can
meaningfully aid legal decision-making about this offender.

However, Table 1 of HMC also includes, for each row, 95\% confidence
intervals for what are described as ``group and individual risks,''
obtained by substituting different quantities into formulae derived
for interval estimation of a proportion by
\citet{WilsonIntervals}. For instance, for the largest (5th) category,
the ranges 0.27-0.44 and 0.03-0.91 are given respectively as 95\%
confidence intervals for ``estimates of risk for groups and
individuals.''  Based on overlaps between the former intervals, HMC
claim that the data support only three rather than nine ``reasonably
distinct group estimates of risk: low (categories 1-4, moderate
(categories 5-7), and high (categories 8-9).'' The usefulness of even
these three categories is then discounted, on the basis that the
widths of the 95\% intervals for ``individual risks'' within each risk
category demonstrate that ``At the individual level, the margins of
error were so high as to render the test results virtually
meaningless.'' Data from 1086 patients stratified by Static-99 scores
into 7 categories are similarly examined with conclusions that, even
with data from almost twice the number of cases, the ``Static-99
yielded only two distinct group estimates of risk: low categories 0-3
and high (categories 4-6+),'' and that the widths of intervals for
individual risks were comparable to those for the VRAG
categories. CM1-3 and HC base similar critiques of ARAIs on widths of
statistical intervals for individual risks obtained from real or
simulated recidivism data.

To appraise this disagreement about the meaningfulness of distinctions
between the nine VRAG categories in Table 1, for which the proportions
of recidivists increase monotonically from 0 to 100\%, requires first
establishing the framework within which statistical terms such as
group and individual risks, risk estimates, and confidence intervals
have established, consensus meanings, and then considering the extent
to which their application to the issue of violence recidivism risk by
CoHaMi adheres to these meanings.

\subsection{Statistical inference}
Methods of frequentist statistical inference, such as invoked by
CoHaMi assume that data available for statistical analysis, such as
the 608 joint observations of VRAG category and recidivism outcome
summarized in Table 1, constitute one possible \emph{realization} of a
data acquisition process that might well have generated different
results from a \emph{sample space} of other possibilities. The sample
space itself reflects an underlying \emph{universe} or, synonymously,
\emph{target population} or \emph{population} for short, that one
wishes to characterize, and from which the data are presumed to have
arisen. Between the sample space of unobserved results and the data we
have actually observed is a mechanism that determines which among the
possibilities in the sample space is revealed to us. The mechanism
itself may be transparent as, for instance, in coin flipping, dial
spinning, selection of lots, or shuffling and dealing of labeled
cards, or opaque to us, as in particle physics at the extreme
micro-level, and with most of the mysteries of human life at the
macro-level. But its behavior is describable by a \emph{probability
  measure} which ascribes numbers between 0 and 1 inclusive, called
probabilities, to many types of characteristics the observed data may
exhibit.

These probabilities are mathematical generalizations of proportions,
i.e., relative frequencies or fractions of times the data exhibit a
characteristic, among all possibilities in the sample space. For most
purposes, including recidivism risk prediction, thinking of them as
simple proportions works well, so long as it is kept in mind that
probabilities represent and reflect, through the sample space, aspects
of the underlying universe and data acquisition process, rather than
of the specific data, i.e., ``sample,'' observed in a particular
instance from that population and process.  Any given sample provides
us with only one of many -- often an infinite number -- of the
possible partial views of the underlying universe we might have
observed.  In such a context, frequentist \emph{statistical inference}
consists of methods for systematically using the data we observe to
make statements about the underlying population and process that have
ascertainable probabilities -- interpreted as long-run relative
frequencies -- of being correct.  The probability that such an
inferential statement is correct is a property of the method used to
create the statement and any assumptions about the population and the
probability measure incorporated into that method.

Quite commonly, we wish to make such statements about numerical
summaries of the population, for instance, the average value of a
measurement, or a proportion with some property (e.g., female), and
the statements we make are based upon its counterpart value or another
numerical summary of the sample. The terms \emph{parameter} and
\emph{statistic} denote any such numerical summaries of, respectively,
a population or a sample. A \emph{confidence interval} is a particular
type of statement made about a parameter, based on one or more
statistics from a sample.  The confidence interval statement claims to
place either a lower bound, an upper bound, or both on the value of
the parameter. Its \emph{confidence coefficient} is the probability,
or relative frequency, with which these bounds are asserted to be
correct.  The concepts of parameter and statistic are both distinct
from any single observation, or \emph{datum}, in the sample.  A
\emph{prediction interval} is a statement analogous to a confidence
interval, but which places bounds on a datum that will subsequently be
observed rather than on a parameter.  Just as the confidence interval
bounds are based on data observed from sampling the population which
the target parameter describes, the prediction interval bounds are
based on previously observed data from sampling the population from
which the new datum will also be produced, usually by the same process
that will produce it.  The linkage of an inferential statement to data
from the population and process targeted for inference is crucial to
the validity of all statistical inference.

The manner in which these concepts apply to the problem of violence
recidivism has been largely left implicit by the recent disputants.
But the following frequentist framework seems compatible with all the
views of all contributors to the current controversy.  Within a
conceptual population of present and future violent offenders
(possibly further delineated by type of offense and sociodemographic
characteristics), people differ in their genetic and environmental
influences, physical and mental attributes, and particular
experiences.  From time to time, circumstances arise with potential to
provoke a violent response.  The frequencies and strengths of such
circumstances vary from person to person, to an extent in relation to
their individual characteristics, and to an extent due to pure
happenstance.  Whether an individual responds violently to any
particular such provocation may depend on specifics of the event, both
large and small, including state of health on that day, presence and
degree of intoxication, and whether there is a weapon immediately at
hand when the provocation occurs.  Because of these factors, whether
or not a violent offender reoffends within a given period is to some
extent random, meaning that we can imagine circumstances in which a
known reoffender might not have done so, and in which an offender who
avoided repeating would instead have been sufficiently provoked to
commit further violence \citep{Appelbaum}.

One way to model such a situation, conceptually and mathematically, is
to presume that i) individuals possess different psychological
thresholds for responding with violent aggression to provocative or
inciting stimuli, so that stimuli whose strengths exceed an
individual's threshold will elicit a violent response, and other
stimuli will not; ii) that these thresholds fluctuate over time for
each individual around a latent mean value, indicative of that
person's general resistance to violence; iii) that provocations of
varying numbers and strengths (whether external such as insults,
perceived threatening behaviors, or actual physical attacks, or
internal such as mental disturbances due to illness or drugs) follow
some form of a random distribution over time, possibly dependent on
the individual's demographic characteristics and aspects of his or her
environment. Within such a framework, an observation of violence
recidivism or restraint on the part of a given individual during a
specified period may be viewed as one realization from among many
alternative ``life scenarios,'' in which provocations of randomly
varying strengths encounter violence resistance thresholds that
themselves fluctuate randomly, but largely within ranges differing
from person to person.

While it is not necessary that all concerned with violence recidivism
risk assessment agree precisely on all details of this framework, this
or some analogous conceptual model is needed to give meaning to the
concept of individual risk as distinct from group risk, and to
consider estimation or prediction of ``individual risk'' using
frequentist statistical methods. While one may conceptually
``individualize'' ideas such as risk by successively considering
groups of individuals specified by increasingly narrow restrictions,
such individualization does not fully accommodate individual
uniqueness which assumes, by definition, that members of any group --
no matter how narrowly defined -- are distinguishable.  We now
consider more specifically the concept of ``risk,'' and the
implications of distinguishing individuals from groups in its
application to statistical inference.

\subsection{Outcomes and risks}

We will use ``outcome'' to denote an individual's experience or
avoidance of a specified event.  This term is used similarly to
``fate'' and ``destiny'' in literature when these refer to ``what
actually happens to someone'' or ``where someone ends up,'' without
the connotations of inevitability carried by the latter from their
origins in Greek and Roman mythology and religious associations. In
statistical usage, ``risk'' is the probability of an undesirable
outcome. From a frequentist perspective this is defined relative to an
appropriate (conceptually infinitely large) population of similar
cases, and can be interpreted as either i) the limit approached by the
fraction of a sample from that population for whom the event occurs,
as that sample is continually enlarged, or ii) the fraction of times
the event occurs within a sample of fixed size, averaged over repeated
random samples of that size from the population.  We can also form
subgroups of the total population, typically defined by shared values
or ranges of specified characteristics, and define \emph{group risk}
similarly.

To accommodate individual uniqueness, a model for occurrences of
undesired events such as described in Section 2.2 would associate
individual risks, distinct from those of others, with each specific
person.  While such individual risks need not be individually observed
or even observable, they jointly constitute group risk which, by the
preceding definition, becomes equivalently the arithmetic mean of its
membersÕ individual risks, or the average individual risk from the
process of randomly sampling a member.  In other words, the group risk
is just the average individual risk of a random member. If we use
$R_{i(G)}$ as shorthand for the risk of the $i^{th}$ member of group
G, then we may use $R_G=\textrm{average}_i R_{i(G)}$ to denote the
``group risk'' for G.

Risks are by definition latent, meaning they inherently characterize
the outcome tendencies of groups or individuals they describe, but are
themselves unobserved. Under certain conditions they can be viewed as
parameters, and estimated or predicted from properly collected
data. But they cannot be fully determined by a particular person's
outcome, or by those of a sample of G's members, since each member's
outcome can vary consistent with his or her individual risk
$R_{i(G)}$.  As a consequence of this, and also of differences in the
memberships of possible samples, the fraction of a sample experiencing
or exhibiting any particular outcome event varies from one sample to
another.

When a population's members are equally likely to be included in a
sample, then the sample fraction experiencing a stipulated outcome
event is an appropriate statistical guess, technically a \emph{point
  estimate}, at that population's group risk $R_G$, and also may
contribute to formation of a confidence interval for $R_G$.  Such a
statistical \emph{estimate} obtained from a sample of a group, say
$\hat{R}_G$ estimating $R_G$, is also commonly denoted as \emph{risk},
while maintaining its distinction from the underlying and still
unknown parameter.  The method of calculating $\hat{R}_G$, in this
case simply averaging the sample, is called an \emph{estimator}; the
estimate is the value the estimator takes when applied to a particular
sample. For instance, the observed $\hat{R}_G=35\%$ proportion of
recidivists in the sample of 116 VRAG Category 5 offenders shown in
Table 1 is a sample risk that might be used to estimate the recidivism
risk $R_G$ of VRAG Category 5 offenders generally.  This sample risk
$\hat{R}_G=35\%$ and its unknown target $R_G$ should not be conflated.
Another sample of 116 offenders would very likely yield a different
fraction $\hat{R}_G$ -- as perhaps would the same selection of
individual members if observed at another time or under slightly
different circumstances, e.g., if an original offense were detected a
few months later, in which case a subsequent event might go undetected
or not occur. $R_G$ is the unobserved but constant overall fraction
describing the population of past, present, and future offenders from
which the samples come.

The distinctions between group risk $R_G$ and individual risks
$R_{i(G)},$ the respective estimates of these $\hat{R}_G$ and
$\hat{R}_{(i)G}$, and the outcome for an individual group member, say
$Y_{(i)G},$ are critically important. It is particularly necessary to
recognize the difference between what might be called
\emph{individualiz\underline{ed} risks} and individual risks. By the
former is meant the group risk for a class of persons sharing similar
relevant characteristics to the person of interest, such as the score
category or even specific score on the VRAG or STATIC-99. An estimate
$\hat{R}_G$ of the group risk $R_G$ for such a collection of
individuals might be judged a more justifiable guess at the individual
risk of each member of such a restricted, homogeneous group than would
be an estimate of the group recidivism risk from a more heterogeneous
group of individuals. If so, an estimate $\hat{R}_G$ of the group risk
of a collection of similar individuals might rationally be applied
uniformly to predict the individual risks of any additional members of
the group who come to the legal system's attention and for whom no
further relevant information is known. The production of such
individualized risk predictions is what ARAIs actually do. The word
``predict'' is used here rather than ``estimate'' because the guess at
an individual's risk is not based on a sample of recidivism outcome
data from that individual, in the manner that $\hat{R}_G$ for $R_G$ is
based on a sample of recidivism outcome data from members of $G$. No
data on recidivism from the new individual, on which a statistical
estimate of $R_{i(G)}$ might be based, have been observed. The
individualized guess, i.e., prediction, of violence recidivism risk is
based upon specified characteristics of the new individual and the
relation of these characteristics to recidivism in the larger, more
heterogenous, class of persons used to construct the ARAI. The
individualized risk predictions of ARAIs are identical for members of
groups sharing the same set of relevant characteristics, and even for
those with different sets of such characteristics that combine to
produce the same ARAI score, or scores in the same ARAI category.

But they are not individual in the sense that persons are unique, and
may have different underlying risks $R_{(i)G}$, even if it is
reasonable to predict these individual risks by the same
individualized estimate $\hat{R}_G$ from the same group $G$ of persons
with similar characteristics. This point is essential to understanding
the root fallacy in HMC, CM1-3's, and HC's treatments of confidence
intervals for individual risks. The statistical intervals that may
legitimately be used for bracketing individualized risk predictions
and individual risks themselves are different, and the latter require
information or assumptions not needed to obtain the
former. Specifically, data must be available or assumptions must be
made about how individual risks vary from one another, from member to
member within the groups from which individualized risk predictions
are formed.

As will be noted below, it is not entirely clear what CoHaMi mean by
the term ``individual risk,'' or that this is used consistently in
these papers. HCM obtain their intervals for individual risks by
adapting an established formula for confidence intervals for group
risks of the aggregated risk score strata in their Tables 1 and 2 (and
our Tables 2 and 3). These are risk estimates individualized only to
the ARAI score ranges defining the respective strata. CM1, CM3, and HC
obtain logistic regression-based intervals corresponding more narrowly
to individuals sharing a single, specified ARAI score. However, CM1
obtain their intervals by adapting the prediction interval for a new
Gaussian observation from a linear regression, and argue that their
intervals do not narrow with increasing sample size, strongly
suggesting that their intent is to provide a range for truly
individual risks.  In contrast, HC's individual risk intervals are
true confidence intervals for group risk of those with a specific ARAI
score. These most certainly do narrow with increasing sample size, and
hence are appropriate for specifying the precision of risk estimates
narrowly individualized to the group sharing a specific ARAI score,
but not for clarifying the range of individual risks within such a
group. Moreover, as will be shown, both HMC and CM1's adaptations of
conventional formulae are mathematically erroneous, while HC
misinterpret the standard logistic regression confidence intervals
they obtain.

Clear discrimination between estimating or predicting any of the types
of risk described above, and predicting ultimate outcomes $Y_{(i)G}$
themselves, is also fundamental to understanding the rhetorical muddle
in discussions of recidivism risk assessment. The outcomes $Y_{(i)G}$
are random occurrences that can only be anticipated, i.e., predicted,
even if individual risks $R_{(i)G}$ are definitively, precisely, known
\emph{a priori}.  For instance, we know \emph{a priori} that the
``risk'' of tails on a coin flip is 50\%, but are totally incapable of
predicting the outcome of each individual flip.  If, however, the
event in question were attempted murder within the next week, we would
presumably at least consider preventive or protective action, despite
this inability to predict whether an attempt will actually occur that
week, and even if the probability of 50\% were not known \emph{a
  priori}, but estimated or predicted with some margin of error.  This
highlights a fundamental problem of risk prediction. The real desired
target of prediction is not the risk, which is a latent intermediate,
but the outcome event itself, which can only be predicted with high
accuracy under two conditions: i) when chance, including the influence
of unobservable predictors, plays little role in determining an
individual's outcome, and thus the $R_{i(G)}$ are close to 0 or 1, and
ii) when these $R_{i(G)}$ can themselves be accurately predicted, so
that persons $i$ for which $R_{i(G)}$ is close to $0$ may be
accurately distinguished from persons $i^\prime$ for which
$R_{i^\prime(G)}$ is close to $1$.  Nevertheless, experience in many
fields, such as medical diagnosis and prognosis, has shown that
prediction need not be highly accurate at the individual level for
major collective benefit to accrue.

\subsection{Statistical intervals}

Confidence intervals were invoked in Section 2.1 to frame and motivate
our discussion, and the widths of confidence intervals and prediction
intervals are at the core of the current controversy. Formally a
confidence interval, sometimes also called an \emph{interval
  estimate}, is a range constructed from a sample with a predetermined
probability of encompassing a targeted population parameter.  The
predetermined probability is the ``confidence coefficient,'' for which
oft-used conventional values are 90 and 95\%.  Such intervals may be
bounded by two statistics from the sample, or by one statistic and a
fixed maximum or minimum possible value for the parameter.  For
instance, risks must be between 0 and 1, so ``one-sided'' intervals
bounded below by 0 or above by 1, or two-sided intervals entirely
within these boundaries, may be useful for different purposes.  A
confidence interval is valid if the confidence coefficient truly
describes the fraction of times the interval will include its target
parameter.  Such validity is a product of the construction of the
interval, that is, the determination of its boundaries, by simple
algebraic manipulation of a probability statement.  Usually this
statement is about a corresponding estimator, i.e., sample statistic,
as described in Section 2.2.2.  The originating probability statement
might give a range in which a function of the estimator will fall for
95\% of samples, with the boundaries of that range expressed in terms
of the target parameter, for instance, $\mu-2 \le \bar{x} \le \mu+2,$
where $\mu$ is the population mean, a parameter, and $\bar{x}$ is the
sample mean, an estimator of the parameter $\mu.$ Simple algebra then
converts this inequality to $\bar{x}-2 \le \mu \le \bar{x}+2,$ after
which substitution of the value of $\bar{x}$, i.e., the estimate, from
the observed sample gives numerical values to the interval's ends.  If
the original inequality is true with probability 95\%, then so is the
equivalent new version.  The probability statement and hence
expression of confidence refer to the proportion of samples for which
the process of interval construction achieves its objective, and hence
for which an inequality with random endpoints is true. The example
says no more than that, if the mean of a sample is known to fall
within two units of the population mean for 95\% of samples, then
whenever that occurs, i.e., in 95\% of samples, the range formed by
moving two units up or down from the sample mean will include the
population mean. Construction of confidence intervals in practice is
more complex, but follows similarly simple logic.

A prediction interval is similar to a confidence interval in also
being a range whose boundaries are random and determined from a sample
of data through a probability statement. But the target of a
prediction interval is not a fixed, unknown parameter characterizing
the population, but rather the value of a random outcome of the
sampling process.  This may be either an unobservable, hence latent,
variable or ``effect'' that describes or governs an aspect of the
sample, or an additional datum not yet observed.  The confidence
coefficient of a prediction interval refers to the fraction of times
the interval will capture its target, whether such a latent effect or
a ``next'' data value.  Since the ``next'' value will be a random
manifestation of the same process that produced the data from which
the prediction interval is obtained, as that process continues to
generate additional observed data values, a 95\% prediction interval
may also be interpreted as an interval designed to encompass 95\% of
such subsequent values.  Because of this, widths of prediction
intervals for additional observations directly portray the extent of
variation between individual sample points, in contrast to confidence
intervals, whose widths reflect only variability of the summary
estimators on which they are based, and thus narrow in inverse
proportion to the square root of the size of the contributing
sample. Thus, for instance, quadrupling the size of a sample may
roughly halve the width of an associated confidence interval, but
leave that of a prediction interval virtually unaffected.

\section{Recidivism risk intervals}
With the background above, we now take up CoHaMI's statistical points
against ARAIs, specifically in turn.
\subsection{Intervals for group risks}
CoHaMi stress that decision-making relying on ARAI-guided recidivism
risk assessments is compromised by statistical unreliability in the
estimation of group risks $R_G$ by within-group recidivism proportions
$\hat{R}_G$.  The argument rests on two grounds:
\begin{enumerate}
\item that 95\% confidence intervals for group risks in ARAI score
  strata, using Wilson's method for interval estimation of single
  proportions \citep[HMC,][]{WilsonIntervals} or intervals based upon
  logistic regression (CM1, HC), are wide and overlapping, generating
  very few genuinely distinguishable risk categories; and
\item that classification error, i.e., potentially fluctuating
  placement of individuals into ARAI score strata due to variability
  in ARAI responses, adds further imprecision, by randomly varying the
  already imprecise ARAI-based group risk estimate $\hat{R}_G$ with
  which an individual defendant is associated (CM1).
\end{enumerate}
Both points are literally true: the 95\% Wilson confidence intervals
do overlap, and assignment of individuals to ARAI-based risk strata
can vary due to error or other sources of variability in an ARAI risk
score, but their contexts and implications have been seriously
misunderstood.

The interpretation of these phenomena is at issue, because ARAI
score-based risk categories and group risks are arbitrary,
instrumental entities, not fundamental targets of interest in their
own right. The choice of a particular categorization is a consequence
of the distribution of risk scores for the particular instrument and
the desire for easily communicated operational guidelines or polices.
The group risks associated with each stratum do not describe fixed,
underlying population subgroups who reside in their corresponding risk
strata; rather, individuals are classified into risk strata based on
responses to ARAIs which, as CM1 stress, are themselves variable.  The
probabilistic process from which the risk estimates are obtained
incorporates two stages.  In the first, individuals who enter the
legal system due to commission of a violent act are deemed eligible
for ARAI-based risk classification and administered the instrument.
Based on the result, each is classified into a risk score category.
In the second stage, each person's subsequent outcome is observed and
used to estimate group risk for that person's risk score category,
presuming that those thus far observed constitute an equally-weighted
random sample of all those who fall in the risk score category.
Chance thus plays a role in determining both the categorization and
the outcome.  Note that this process does not assume that individuals
arise from underlying ``true'' risk categories which their ARAI scores
imperfectly reflect, although some ARAIs may be compatible with such a
assumption.  Rather, the meaning of the risk categories used for
prediction derives directly and only from single observed and
impermanent responses of individuals to the ARAI. The group risk
parameter $R_G$ which the observed risk $\hat{R}_G$ estimates in such
a situation is then the long-run fraction who return to violence among
those whose scores on their diagnostic ARAI administrations fall in
the relevant risk score category, regardless of the fraction of these
who may have been in some sense misclassified due to misreporting or
other types of error.

Although it seems attractive to estimate such risks precisely, the
justification for this intuition is weak.  Such an $R_G$ need not be
the individual risk $R_i(G)$ of any single member of its ARAI score
stratum, nor does it characterize the average risk of any uniquely
identifiable subgroup of violent offenders, since the membership of
each risk stratum is not inherent to the members but arises to some
extent randomly.  Rather, it is a statistical property of the
assessment process.  Moreover, even if known precisely, the collection
of $R_G$ for all ARAI-based risk score strata is insufficient to
describe the effectiveness of a risk assessment procedure without the
additional knowledge of the frequencies with which offenders fall into
each of the respective categories.  And even were both these types of
information known with perfect accuracy, further mathematical
manipulation would be required to obtain simple measures of prediction
model benefit that reflect the influence of a risk prediction and
management policy on actual outcomes, e.g., the fraction of recidivist
violence that could be prevented by incarcerating offenders above a
stipulated ARAI threshhold.  Note that the evaluation and validation
of prediction model performance has been the subject of extensive
statistical research, yielding numerous direct methods unrecognized by
and superseding these authors' confidence interval approach
\citep{HarrellLeeMark,Harrell,Pepe,Steyerberg,ZhouObuchowskiMcClish}.

If focus on variability of the $\hat{R}_G$ and their linkage to
individuals is nevertheless considered important, however, the
objections (1) and (2) above are both readily seen to be
misleading. Regarding (2), variability in assessment and reporting of
the characteristics contributing to an ARAI, and hence in ARAI scores
and classifications based upon them, is as present in the examinations
used to construct and validate an ARAI algorithm as in applications of
ARAIs to subsequent offenders. Associations of ARAI scores and score
categories with subsequent recidivism are based on data subject to
measurement and misclassification error, fallible as these data may
have been. If some idealized form of the ARAI from which such
variability could be removed were to become available, such
associations would likely become stronger. But, for now, while such
variability must be acknowledged, its impact cannot be regarded as a
previously unrecognized defect superimposed upon the ARAI from outside
and degrading its performance relative to prior expectations. The
effects of such variability are already reflected in data such as in
HMC's Tables 1 and 2. Moreover, the existence of such variability and
misclassification in itself is no issue, since measurement and
diagnostic variability pervade virtually all aspects of clinical
medicine and psychology that have been closely examined. CM1's basic
point that ARAI classifications can vary applies as well to virtually
all other clinical classifications and judgments. That the explicit
quantitative nature of ARAI-based assessments allows explicit analysis
and worrisome conjectures about the effects of measurement variability
and misclassification is a virtue rather than a defect of ARAIs, and
does not suggest that more qualitative alternatives less subject to
explicit analysis are in any sense superior.

With respect to the precision of individualized risk estimates for
ARAI score categories, narrower confidence intervals than those found
wanting by these authors are readily available from the current
data. Such intervals might be further narrowed by incorporating larger
samples, without need to question the general enterprise of ARAI-aided
risk assessment.  For instance, HMC take an unusually conservative
approach to interval estimation for circumstances where expectation
and evidence both strongly support monotonic (steadily increasing)
$R_G$ with increasing ARAI scores. Their conservatism takes two forms:
the determination of each interval in isolation from the information
about trend provided by data from other risk categories, and the high
95\% confidence coefficient required for each separate interval rather
than, perhaps, for an underlying trend parameter.  CoHaMi argue
strongly for logistic regression modeling as the proper source of
inference for data from ARAIs.  While categorization introduces
measurement error into statistical models and use of original scores
for each individual is usually preferable, models based on reasonable
quantitative scores for ordinal categories may be useful for both
description and inference. (Such models will usually tend to
underestimate the predictive power and precision that can be obtained
from alternative models based on raw rather than categorized data, and
hence will tend to understate the discriminating power of an ARAI.)
In this spirit, Figures 1a and 1b summarize fits of simple logistic
regression models to HMC's Tables 1 and 2, presuming equal spacing of
the ARAI categories on the usual logistic regression scale.  Each
figure plots $\hat{R}_G$ on the vertical axis against the sequence
number of the ARAI risk categories, from low to high risk.

\begin{center} FIGURES 1a AND 1b HERE \end{center}

\begin{center} TABLE 2 HERE \end{center}

\begin{center} TABLE 3 HERE \end{center}

Tables 2 and 3 respectively compare the VRAG and STATIC-99 95\%
confidence intervals shown in these figures with those of HMC, showing
considerable narrowing and reduced overlap of intervals from both
sources. By HMC's criterion there are five distinguishable risk
categories for VRAG and four for STATIC-99, as compared to three and
two respectively found by HMC.  However since, as noted above, 95\%
confidence coefficients for wholly disassociated intervals is a
decidedly stringent approach that produces wider intervals than would
be required, for instance, to test the hypothesis of equal risk
between two ARAI score categories, many would consider it reasonable
to relax these intervals to a less stringent 80\% confidence level.
(Note that the conventional use of 95\% intervals stems from the
perpetuation of a ``convenient'' choice by R.A. Fisher in using
confidence intervals for a different purpose
\citep[p. 47]{FisherSMRW}.) Hence, Tables 2 and 3 also include 80\%
confidence intervals. These show no overlapping categories other than
a 0.2\% overlap at the boundary of the two lowest VRAG categories, the
bottom one containing only 11 offenders. SAS 9.3 code for Figure 1 and
the logistic regression intervals in Tables 2 and 3 is appendicized.

Any Wilson or logistic regression-based confidence interval, whether
pertaining to an ARAI score category or raw ARAI score, can for any
chosen confidence coefficient be narrowed by enlarging the sample from
which it is inferred.  A given interval will not invariably be
narrowed by expansion of any possible sample, but such narrowing is
likely and mathematically inevitable with sufficient enlargement.
Thus, intervals based on enough data will eventually fail to overlap
unless they narrow to the same overall risk $R_G$.  While the latter
is possible in principle, the available data exhibit very highly
statistically significant trends toward increased recidivism with
increasing VRAG and Static-99 score categories, specifically
$P<0.0001$ for both logistic regression models above. The
relationships of these ARAIs to violence recidivism in each existing
data set would have to be remarkably non-representative of the
corresponding relationships in the populations and environments from
which they stem for such neighboring confidence intervals to so
converge.

\subsection{Intervals for individual risks}
In parallel with the confidence intervals for group risk, HMC present
95\% confidence intervals for individual risks of members of each of
the nine VRAG categories based on what is described as an \emph{ad
  hoc} use of Wilson's confidence interval formula. HMC interpret
these intervals as meaning that ``Given an individual with an ARAI
score in this particular category, we can state with 95\% certainty
that the probability he will recidivate lies between the upper and
lower limit.''  The stated justification for this use of Wilson's
confidence interval formula is as a heuristic approximation to
logistic regression results. In response to criticism of such use of
Wilson's formula (\citealt{MossmanSellke},
\citealt{HartMichieCookeReply1}), CM1 provides intervals based on a
different formula, claimed to be directly based on logistic
regression.  In this section, we first address why neither valid
confidence intervals nor valid prediction intervals for truly
individual risk can ever, in principle, be obtained from data such as
used by CoHaMi, using any method.  We then indicate specifically the
respects in which the formulae used by HMC and CM1 have been
mathematically misapplied, as well as how HC have misconstrued and
misinterpreted intervals for individualized group risk as pertaining
to truly individual risk.

Recall that confidence intervals and any other form of frequentist
statistical inference proceed, by definition, from probability
statements about data that have been sampled from the target
population or process about which inference is to be made.  Confidence
intervals for group risk proceed from probability statements about
recidivism that has been observed in samples of offenders who are
first classified into strata on the basis of the ARAI scores.  They
pertain to individuals whose ARAI scores place them in the
corresponding group.  HMC create a Wilson confidence interval for
group risk for each such stratum from a single row of Table 1, i.e.,
using data from that stratum.  The logistic regression approach above
makes an assumption about the form of the relationship between the
group risks and the score strata, allowing a mathematical stitching
together of information from the different strata to produce narrower
intervals.  Both types of intervals arise from standard methods of
statistical inference, in which data from the process of ascertaining
a sample of offenders from a defined population, classifying the
offenders using the ARAI, and monitoring these offenders for
recidivism, are used to form confidence intervals for parameters $R_G$
of the process which generated those data.  Statistical inference is
justified by this basic relationship: data from a process are used for
inference about that process, as described in Section 2.2.

These group intervals estimate the mean latent risk $R_G$ among all
offenders classified into the same risk stratum.  If all individuals
in such a stratum were to have identical risk by virtue of their
stratum membership, then ``individual risk'' would have no meaning
distinct from the stratum's group risk.  But clearly CoHaMi, in
distinguishing individual from individualized group risks of their
corresponding ARAI score strata, are trying to go further in targeting
and attempting to describe the variation of individual offenders among
those in the the same ARAI score stratum.

How might one distinguish different levels of risk, say $R_{i(G)}$,
for otherwise presumptively similar individual members of the same
group $G$ identified by an index $i$ (which might represent simply
their order in a list)?  This is relatively easy, at least
conceptually, when the risk pertains to episodic events for which the
individual is repeatedly observed, and when risk is stable across
observations.  In health care, new dental caries, migraine headaches,
epileptic seizures, multiple sclerosis relapses, and incontinence
episodes are events for which individuals might be repeatedly observed
over discrete periods, with accumulation of the fractions of such
periods during which an event of concern occurs.  High risk
individuals experience events in more periods than those at low risk
and, in periods when the disease process is stable, prognosis may be
estimated from past experience.  In such circumstances, the individual
becomes a system from which data are observed, and a person's risk is
inferable from data on that person's own experiences.  The target of
inference is a parameter of the process generating a single person's
data, and inference is made from that single person's data to the
unique individual risk parameter from which that person's data arose.
We emphasize that data and inferential target remain linked, but now
within a single person rather than a single ARAI score stratum.

Assumptions about distributions of risk across individuals, and
relationships among the experiences through which the underlying risks
of different persons play out, may also be used to stitch together
such clusters of intra-person observations (technically, ``repeated
measures'' data) across groups of persons from one or many ARAI
strata.  For instance, in a ``mixed effects logistic regression
model," counts of periods in which pertinent events are observed from
each of a set of individuals, each of whom has been classified into
one of several strata and observed for multiple periods, are assumed
to each arise from a process behaving like independent flips of a
weighted coin, with the process in one individual unaffected by that
in any other.  A similar assumption as in logistic regression is made
about the relationship of risk to ARAI score stratum, and a simple
function of risk is assumed to vary randomly across individuals within
each score stratum according to a normal (Gaussian) probability law,
the well-known ``bell-shaped curve.''  This overall framework,
including sampling of individuals, classification into strata, the
numerical scale and Gaussian nature of variation from one individual
to another within the same stratum, and the relationship of risk to
stratum, allows estimation from a sample of this process of i)
inter-stratum risk trends, ii) means and variances of the normal
distributions underlying intra-stratum individual variation, and iii)
prediction intervals for individual risks.

CM1's and HC's attempted confidence intervals for individual risks are
analogous to this latter aspect of modeling repeated measures binary
data using mixed logistic regression
\citep{FDVM,GeertGeert2005,Vonesh2012}, but differ in that elements of
the preceding formulation are absent.  These missing elements are: a)
recidivism outcome data from offenders from whom individual risk
intervals are desired, b) repeated recidivism outcome data from any
individuals, and c) assumptions or information about variation of
individual risk among offenders within risk strata or with identical
risk scores.  Absence of these elements makes it impossible to create
statistical confidence or prediction intervals for individual risk, in
principle, by any methods within the realm of frequentist statistical
inference.

Here is why.  To reiterate, frequentist inference means inference from
data generated by a process about aspects of that same process.  These
aspects may be parameters that describe and govern the process, with
confidence interval estimation being one type of statistical inference
about such parameters.  They may also be random results of the process
that have yet to be observed, or have already occurred but are
unobservable directly but indirectly inferable through subsequent
data.  Prediction intervals are a form of statistical inference about
such latent variables or future observations.  In any case, logically,
unless one observes data on the recidivism process specifically from a
new offender (a), statistical inference about that new offender's
individual risk must be based on assumptions linking the new offender
to a population/process from which recidivist outcome data are
available to serve as the source of inference.  Assumptions and/or
data on the distribution of individual risk within that
population/process (c) are also needed, to show how data from others
can actually inform prediction of the new offender's \emph{individual}
recidivism risk.  Since CM neither articulate nor refer to such
assumptions, the needed information for their individual intervals can
only have been data-based: obtained from data on variation of
individual risks among offenders sharing a common risk stratum, whose
recidivism outcomes have been observed.

This, however, is impossible.  The data used by CM1 consist of PCL-R
score and a single binary recidivism outcome, reconviction with prison
sentence for violence, for each of 255 offenders.  Data such as these,
where a binary outcome is measured only once on each individual,
provide no information to distinguish variation among individual risks
from random variation of outcomes among those whose outcomes are not
inevitable, that is, for whom $0 < R_{i(G)}<1$ .  An example makes
this clear.  Consider a sample of size two, from either of two
scenarios, where the population consists of a single very large
stratum within which either A) all individuals have risk
$R_{i(G)}=60\%$, or B) each individual's outcome is inevitable: 60\%
are inexorably destined to recidivate, i.e. have $R_{i(G)}=100\%$,
while the other 40\% are either incapacitated or totally resistant to
violence, and will never do so, i.e. have $R_{i(G)}=0$. These
scenarios describe diametrically opposed individual risk profiles. In
A), all individuals have exactly the same propensity to recidivate
$R_{(i)G}$, and there is great uncertainty about each individual's
outcome. Another way of describing this is that all variation in
outcomes arises from chance variation in the experiences of each
individual rather than from differences between individuals. In B), on
the other hand, the $R_{(i)G}$ vary dramatically, and all variation in
outcomes arises from which individuals are included in a particular
sample.

Now suppose also that, in each scenario, members are chosen with equal
probabilities, with each choice of a member of either population made
independently of the others, and the recidivism outcomes of each
sampled member then observed.  In A), the two members chosen will have
identical 60\% risks, and the probabilities of two, one, and zero
recidivists among them are easily seen to respectively be
$0.6\times0.6=0.36, \;0.6\times0.4+0.4\times0.6=0.48, \; \textrm{and}
\; 0.4\times0.4=0.16$.  In B), each member's outcome is predetermined,
but the probabilities that the sample will contain two, one, or no
members destined to be recidivists will have the same values as above,
by the identical numerical calculations.  Whether the random component
is in the sampling process and attributable entirely to variation
among essentially inevitable individual destinies, or in the outcome
process and attributable entirely to random experiences and responses
unique to each observation, and independent of the individual, has no
effect on the distribution of possible observations.  Since the
probabilities of possible samples are totally insensitive to whether
the data arise from a distribution of distinct predetermined outcomes
-- the most extreme case of highly individual risks -- or from a
single value of risk shared by all individuals in common, the data
cannot inform a choice between the two scenarios, and thus provide no
information about distribution of individual risks.

Put another way, without repeated measurement there can be no
observable basis for differentiating between these two diametrically
opposite characterizations of individual risks. While truly individual
risk may remain an instrumentally useful mental construct in such
situations, attempting to actually estimate such risks, \emph{when
  they have no real world manifestations}, seems a fruitless exercise
in reification.

This example is essentially totally generalizable.  Information
pertaining to variation in individual risks of binary outcomes is
obtainable only from data on repeated observations of multiple
individuals, and not from single observations on each, no matter how
many individuals may provide them.  If, in Scenario A), we were to
make five repeated observations on each of ten individuals, the
extreme variation in individual risks would be evident from the total
consistency of responses within each of the individuals sampled.  On
the other hand, five repeated observations on ten individuals in
Scenario B) would likely show mixed results in most or all subjects,
following a pattern of variability compatible with results of flipping
the same 60-40 weighted coin five times, counting the numbers of heads
and tails, and repeating this entire process nine additional times.
Such results are easily explained without requiring variation of
individual risks from the overall group risk.  It is only the
clustering of like results among individuals, rather than the actual
results themselves, that provides information about the presence and
extent of truly individual risks.  If data are not collected in such a
way that the extent of such clustering is observed or observable, then
statistical inference about individual risk is not possible from those
data.

The nature of the repeated observations required to apply formal
statistical inference techniques to estimate a truly individual risk,
that is, risk conceived of as idiosyncratic and potentially divergent
from any patterns observable in others, is conveyed by reference to
\citealt{Mulvey} and \citealt{Odgers}. These researchers conducted
separate weekly behavioral interviews with each member of a sample of
132 mentally ill individuals at high risk for frequent involvement, as
well as ``collateral informants'' of each. The interviews revisited
the past week's events in eight domains of life, including violence
and involvement in the mental health treatment and legal systems,
whenever possible at the daily level. Formal statistical methods might
be applied to data of this sort to distinguish between individuals in
the same ARAI risk category who nevertheless, for reasons not captured
by the ARAI and perhaps not systematically ascertainable, have
differed and hence might be expected to differ in the future in the
frequency of violent behavior. Note that this kind of statistical
inference, as all statistical inference, requires data from the system
or process at which inference is directed. Here the systems and
processes are the specific individuals involved. However, this type of
sampling of individual behavior is clearly not applicable to the
circumstances in which ARAIs are needed and employed. We will return
to the general implications of this point in Section 4.

\subsection{Misadventures in individual risk estimation from actuarial data}
The VRAG and Static-99 data from which HMC derive individual intervals
based on Wilson's method, the RM2000/S data from which CM3 make a
slightly weaker claim based on a sequence of intervals generated
similarly, and the PCL-R data from which CM1 derive individual
intervals based on logistic regression, each contain only single
recidivism outcomes on each offender.  The preceding argument implies
these intervals must be statistically incorrect.  This section
identifies where the rationales of HMC, CM1, and CM3 fail.

\subsubsection{Individual intervals from Wilson's formula}
It is helpful to grasp the basic components and characteristics of
Wilson's confidence interval formula. From HMC, the interval's bounds
are
\begin{equation}
  \frac{\hat{\theta}+\frac{z_{a/2}^2}{2n} \pm z_{a/2} \sqrt{\frac{\hat{\theta} (1-\hat{\theta})}{n}+\frac{z_{a/2}^2}{4n^2}}}{1+\frac{z_{a/2}^2}{n} }.
\end{equation}
where $n$ is the number of individuals whose dichotomous outcomes have
been observed; $\hat{\theta}$ is the observed risk, e.g., the fraction
for whom recidivism is observed ($\hat{R}_G$ in Section 2.3); and
$z_{\alpha/2}$ is a mathematically determined value, technically the
number exceeded by $100 \times \frac{\alpha}{2}\%$ of random
observations from a normal distribution with a mean of $0$ and
standard deviation of $1$.  As $\alpha$ decreases, $z_{\alpha/2}$
increases and the interval widens, raising the chance of including the
true risk ($R_G$ in our and $\theta$ in HMC's notation, both hatless).
An interval constructed using $z_{\alpha/2}$ has an approximately $100
\times (1-\alpha)\%$ chance of including $R_G=\theta$ in circumstances
compatible with the assumptions under which it was mathematically
derived.  HMC use $\alpha=0.05$, for which $z_{\alpha/2}=1.96$, to
obtain the $95\%$ intervals for group risks shown in Tables 2 and 3.

The relationship of $z_{\alpha/2}$ to the confidence coefficient,
i.e., the chance that the interval will include its target parameter,
is based upon one of the oldest and most consequential results of
probability theory, Laplace's Central Limit Theorem.  This
mathematical approximation theorem describes the random behavior of
counts and proportions resulting from $n$ independent, identical coin
flips -- or any other phenomenon that may be modeled by them -- as the
number of flips $n$ increases.  Laplace's result, that the binomial
probability measure governing such counts and proportions may be
approximated by a normal probability law as $n$ increases, implies a
probability statement about the random behavior of observed risks
$\hat{R}_G$ as the samples from which they are derived increase in
size.  This statement is manipulated to produce the Wilson interval
bounds in equation (1), and embodies the relationship of the
confidence coefficient to $z_{\alpha/2}$ on which the validity of
Wilson intervals thus relies.  Laplace's Central Limit Theorem itself,
and hence this probability statement validating the use of Wilson
intervals, depends entirely on the observed proportion $\hat{\theta}$
representing a sample of precisely $n$ individuals.  The manipulation
of the probability statement to obtain the confidence interval bounds
given by equation (1) is valid only when this sample arises from the
group whose true proportion, here group risk, is the inferential
target.  Moreover, the mathematical approximation in Laplace's Central
Limit Theorem is inaccurate when $n$ is small.

To show HMC's and CM3's use of (1), Table 4's first two rows adjoin
material from HMC Table 2, based on a sample of offenders scoring 0 on
the Static-99, with material from the middle column of CM3 Table 1.1,
based on an RM2000/S medium risk sample.  The percentages of
reoffenders are each 13\% within rounding error.  The intervals, from
HMC and CM3 after correcting CM3's last-digit misprint, are correctly
described by both papers as pertaining to group risk, which we have
called ``individualized'': using a reference group from the offender's
ARAI risk score category calibrates risk to offender characteristics
captured by the ARAI, although the risk is estimated from recidivism
data of others.  The next row shows, from a hypothetical 100,000
offenders of whom 13\% also reoffend, how a large sample can increase
the precision with which such an individualized risk is assessed, by
narrowing the confidence interval to any desired degree.  The
rightmost column distinguishes this sample's hypothetical nature from
the actuality of the prior rows, using the term ``Observable'' to
further indicate that such a large offender sample, if it were
practical to obtain, could actually yield 13,000 reoffenders and
validly generate the narrow 12.8\%-13.2\% interval.

\begin{center} TABLE 4 HERE \end{center}

HMC and CM3 invoke hypothetical scenarios, in which successively
smaller actuarial samples exhibit the same 13\% observed proportion
reoffending, to make this argument in the other direction.  To
demonstrate that confidence intervals widen as sample size decreases,
intervals are obtained by holding $\hat{\theta}=13\%$ constant in (1),
while successively inserting $n=50, 10, 5,$ and finally $n=1.$ These
scenarios and intervals, at the bottom of Table 4, are termed
`Fictitious'' as they require splitting persons into recidivist and
non-recidivist portions, although encounters with a disembodied arm or
leg beating a lover, while the rest of the attacker reads the morning
paper over coffee, are confined to cartoons, dreams, and written or
cinematic horror fiction.  Nevertheless, HMC interpret the interval
for $n=1$ as a confidence interval for individual risk, our
$R_{(i)G}$, for an offender in the group with STATIC-99 score 0.  CM3
interpret the progressive widening of this sequence of intervals as
indicative of ``large margins of error for estimates of likelihood of
individual cases.''

These interpretations are technically incorrect for several reasons.
\begin{itemize}
\item Numerical results from inserting fictitious scenarios into
  Wilson's formula are not confidence intervals.  The statistical
  meaning of Wilson's formula derives from a mathematical model in
  which numbers represent events potentially observable in the real
  world. This is a major issue of mathematical definition and logic,
  not a minor one of numerical rounding or choice of
  terms. ``Statistics'' is a science of description and inference from
  observation.  Formulae and numbers without such linkage to a data
  collection process are outside the province of Statistics and do not
  follow its rules.  ``Confidence intervals'' based on such fictitious
  inputs to Wilson's formula are numerological rather than
  statistical.
\item Even for observable data, the approximation on which Wilson's
  formula is based fails for very small $n$, so resulting intervals
  even from observable data are not valid.  For instance, when $n=1$
  and true risk is 20\%, the supposed 95\% confidence interval
  includes the true risk only 80\% of the time.
\item Most tellingly, changing $n$ cannot in and of itself alter the
  target parameter of a statistical inference.  Hence, regardless of
  $n$, Wilson's formula applied to group data can't have implications
  for a distinctively individual risk $R_{i(G)}$.  It is tautological
  that $R_{i(G)}$ can't be differentiated from $R_G$ by recidivism
  data solely from others.  This means not that a statistical interval
  is indefinitely wide, but that a \emph{statistical} interval for a
  \emph{truly individual} recidivism risk can't exist without
  recidivism data from the individual.
\end{itemize}
Parenthetically, in addition, serial recidivism data on one or more
individuals would likely exhibit dependence patterns, with
observations close in time more associated than those further
separated.  If so, Wilson's formula would not be appropriate even for
use with serial data on one or more individuals.

\subsubsection{Individual intervals from logistic regression formulae}
CoHaMi argue that the Wilson-based intervals adequately approximate
more formally correct intervals available from logistic regression
models for raw ARAI data.  CM1 base these intervals on standard
prediction interval bounds from linear regression \citep{CookeMichie,
  CookeMichieCorrig},
\begin{equation}
  B_0+B_1x_{n+1} \pm t_n \sqrt{\hat{\sigma}^2 \left ( 1+\frac{1}{n}+\frac{(x_{n+1}-\bar{x})^2}{SS(X)} \right )},
\end{equation}
which they convert into probabilities using the logistic
transformation (CM1),
\begin{equation}
  Pr(\textrm{event})=\frac{1}{(1+e^{-z})}
\end{equation}
In equation (2), $B_0$ and $B_1$ respectively represent the estimated
slope and intercept from a standard simple linear regression model
based on $n$ observations; $\bar{x}$ is the average in the sample of
the single predictor $X$, here the ARAI risk score; $SS(X)$ is the
``corrected sum of squares'' $\sum_{i=1}^n (x_i-\bar{x})^2$ of this
predictor, i.e. these ARAI scores; $\hat{\sigma}$ is the estimated
standard deviation of individual outcomes of those with the same value
of the predictor $X$ around the mean response (height of the
regression line) for that specific value $x$ of the variable $X$ of
the predictor, i.e., specific ARAI score; and $t_n$, which also must
depend on $\alpha$ and is generally indexed as $t_{n;1-\alpha/2}$,
plays an analogous role to $z_{1-\alpha/2}$ in equation (1)
\citep{Student}. Writing $Z=B_0+B_1x$, CM1 explain this material
correctly, but then err in claiming ``We have a linear regression of Z
on x so the equation for the CI for Z is the same as the linear
regression case,'' thereby justifying construction of claimed
confidence intervals for individual risks obtained through equation
(3).  Although the context is different, the technical error here is
of the same nature as occurred in HMC's construction of the
Wilson-like intervals: a failure to recognize the dependence of
statistical formulae on their probabilistic bases.

Specifically, in simple linear regression $Z=B_0+B_1x$ is the
estimated mean of continuous observations sharing a stipulated value
of a predictor, and which vary around their average for that value
according to a normal (Gaussian) distribution with mean $0$ and a
common standard deviation, $\sigma,$ across all values of the
predictor.  Under this model, individual observations sharing a given
predictor value may themselves take any numerical value whatsoever. It
is then perfectly reasonable to develop a confidence interval for the
mean at a given predictor value, or a prediction interval for an
unknown continuous observation, which may in principle take any value
in that interval or even outside it.  However, the model presumes that
average values of any individuals with the same predictor value are
identical; thus, there is no distinction between group means and
individual means.  Such a distinction can be introduced, and the model
elaborated to accommodate it, but only in the context of repeated
measurements of individuals.  As noted above, without such repeated
measurements there is no information to distinguish within from
between-individual variation, and hence no information from which to
infer the existence of individual means distinct from group means, nor
of individual risks beyond their individualization based on the
differing ARAI values represented by $x$.

In logistic regression, on the other hand, $Z$ represents the
logarithm of the odds $O=Risk/(1-Risk)$ of occurrence of a binary
event, which can only occur or not occur.  There is no continuous
distribution of data values around either $Z$ or the logistic
regression line obtained by plotting (3) against the predictor value,
here the precise ARAI score, $x$.  Outcomes are binary with values 0
and 1, their means being the probabilities estimated by substituting
$Z$ into equation (3), but with standard deviations functionally
dependent upon and determined entirely by these means. The parameter
$\sigma$ and its estimate as represented by the symbol $\hat{\sigma}$
in equation (3) have no meaning in this context, because variability
is not constant and depends on the value of $x$, the ARAI score.
Equation (2), while highly useful for linear regression, does not
provide a valid confidence interval in the logistic regression setting
because its probabilistic justification is absent in logistic
regression, where the use of intervals to predict individual
dichotomous classifications is also fundamentally misconceived.
Although values $t_{n;1-\alpha/2}$ do indeed appear in correct
formulations of confidence intervals from logistic regression models,
the manner of and justification for their use differ from CM1's, and
the intervals for which they are used do not target individual risks.
As with HMC's Wilson-like intervals, CM1's claimed logistic regression
intervals for individual risks lack foundation in statistical science.

Lastly we consider HC, to whom a debt of gratitude is owed for
publishing, in their Table 2, the specific code in the language of the
Stata/SE statistical software package with which their ``individual
risk estimates and margins of error'' were obtained.  HC use a data
set consisting of 90 Canadian subjects convicted of sexually-related
crimes and followed for an average of 4.2 years, of whom 16 were
categorized as having ``failed'' by virtue of an additional
``investigation, charge, or conviction for a sexual offense or
sexually motivated offense.''  Logistic regression was used to fit
this dichotomous outcome to scores on four domains of the
SVR-20. Subjects were then scored by the results of this fitting
procedure, and partitioned into the highest third and lowest
two-thirds. Confidence intervals were then obtained for the risks
$R_g$ of each group, from the corresponding observed risks
$\hat{R}_G,$ using a conventional, approximate method: $\hat{R}_G \pm
z_{1-\alpha/2} \sqrt{\hat{R}_G(1-\hat{R}_G)/n}.$ These were noted to
be wide but only barely overlap, and HC recognize that these widths
could be narrowed and overlap avoided by forming the model from a
larger group of subjects.  This suggests the uncontroversial point
that ARAIs, to provide reliable estimates of group risks, should be
constructed with much larger samples than 90 subjects.  Although the
categories within which group risks are described were formed using a
logistic regression modeling process, HC's method of estimating group
risk by the simple observed proportion is a common approach that is
not itself based on either Wilson's formula or logistic regression.

HC then examine ``individual risk estimates and margins of error.''
The published STATA/IC code shows that their risk estimates are the
predicted individualized risks for specific ARAI scores, and that
these and the associated error margins were produced as were the lines
corresponding to predicted probabilities and their 95\% confidence
intervals portrayed in Figure 1 above, though using different data and
a different software package. These intervals must thus be interpreted
as plausible ranges for the group risks of collections of individuals
sharing a particular ARAI score, and not as intervals expected to
encompass any given fraction of true individual risks, if these were
to vary within such a group. That they are wide, as HC note, is hardly
surprising however, because they are derived from fitting a four
variable logistic regression model to data from 90 subjects including
only 16 recidivists, in contrast to the VRAG data on 618 subjects
including 192 recidivists, and the STATIC-99 data on 1086 subjects
including 272 recidivists, analyzed by HMC and revisited above. That
ARAIs based on few offenders can give unreliable results is
indisputable, but this does not constitute a critique of the ARAI
project. Concern would be warranted, as CM1 express, if individualized
intervals obtained in this or a similar manner could not be narrowed,
and hence ARAI-based individualized risk estimates made more precise,
by incorporating data from more offenders in their production. To
resolve the question of whether such logistic regression-based
individualized intervals as HC's narrow with increasing sample size,
one may rerun the STATA/IC code after appending the characters
``[fweight=10]'' to HC's first line of code, labeled as Step 1 in
their Table 2, and then rerun using [fweight=100], [fweight=1000],
etc..  These tell STATA to first expand the sample by a factor of 10
to be more in line with the preceding VRAF and STATIC-99 data, and to
then expand it further.

\section{Summary and Conclusions}
In the frequentist tradition of statistical inference from which
CoHaMi attempt to draw, individual risk is an unobservable latent
probability, to which the recidivism outcome is linked as the
realization of an underlying random process. In the absence of further
assumptions, frequentist inference about the risk of a specific
individual can only be made by observing outcomes of that specific
individual's unimpeded underlying random process, repeatedly. This is
neither possible nor desirable for assessing violence recidivism. So
assumptions must be made.  If individual risks were to vary
deterministically with combinations of known, discrete, measurable
factors, then persons might be grouped by combinations of these
factors into internally homogeneous strata, with members of each
sharing the same individual risk, but these shared risks varying among
strata.  Such circumstances would allow statistical inference about a
specific individual's risk from observations of a series of outcomes
of others with the same combination of determining factors, and thus
in the same risk stratum.  Moreover, a model for how risk varies among
some but not all combinations of determining factors might be
interpolated or extrapolated, with outcomes from some strata used to
estimate risks for other strata for which members' outcomes are
unavailable, assuming the model applies. Individual risk in such
circumstances would be synonymous with the group risk of the
individual's risk stratum.

This is the situation ARAIs emulate. But in reality some risk
determinants and correlates will always be unknown, unmeasurable or
imperfectly measured, or vary continuously.  Individual risks may be
expected to be more homogeneous within a risk stratum than in the
general population, but will vary to some degree. Without repeated
observations of outcomes, or further knowledge or assumptions about
how this remaining variation arises, frequentist inference about
specific latent individual risks, or even their spread, is impossible
because the sampling behavior of data depends exclusively on the
underlying population or stratum's mean (i.e., group) risk, regardless
of the risk homogeneity or diversity of its members (Section 3.2).

More explicitly, actuarial risks are mean risks, defined based on
strata and/or models and applied to a new offender using outcome data
from other members of the same stratum or with similar model
score. Although initiated by an ARAI-based index of an individual's
specific characteristics, the index's components, their relative
weights, and the actuarial risk itself stem entirely from the
recidivism experience of the ARAI's reference group. Such risks are
thus better termed \emph{individualized} rather than
individual. Whether broadly or narrowly individualized, which depends
on the ARAI components and how finely their distributions are
partitioned to form strata, actuarial risks \underline{are all group
  risks}.

While individualized (actuarial) risks are produced by a process of
sampling and observation of a reference group that is observable and
amenable to statistical treatment, such treatment does not and is not
capable of describing variation between latent risks conceptually
unique to individuals. The issue is not that individual risks are
exceptionally variable, or statistical intervals of any sort
inherently too wide, or statistical methods lacking.  It is that no
relevant data are available to address the question as framed.  If the
target of inference is individual risk distinct from what is captured
by the components of a comprehensive ARAI, then the individual's ARAI
data are irrelevant to the question as framed.  Unless outcome data
are available from the individual, there are no relevant data on the
target of inference from which statistical inference might proceed.
Statistical inference is excluded essentially tautologically, by
definition.

Much confusion about individual risk assessment is thus of semantic
origin, due to a subject/object mixup.  Phrases such as ``her risk''
are conveniently brief. But by placing ownership of risk with the
offender on whom it is projected rather than the reference group from
which it was derived, such phrasing conflates the concepts of
individualized risk, derived from an external group and projected onto
the subject, and latent individual risk intrinsic to the subject
herself. This invites confusion between statistically assessable
variability in the ARAI production process and variability of
conceptual, latent quantities, intrinsic to individuals, for which
relevant data are unavailable and hence statistical assessment
inapplicable.

This is the trap into which Cooke, Hart, and Michie have stumbled.
What troubles them is the inability of ARAI's to narrowly
statistically bound the latent, unobserved individual risk that an
individual, with all the idiosyncrasies of human variation, may
harbor.  Their papers repeatedly stress imprecision of ARAI-based risk
attributions to individual offenders, believed inherent in ARAIs, as a
primary reason for abandoning the ARAI approach.  Indeed, if tight
statistical intervals for truly individual risks are considered
essential, then actuarial and all other risk assessment must be
abandoned because there is no conceivable way to provide them.  But
this whole project misconceives the nature of actuarial risk
estimation and the source of its espoused benefits. In principle,
precise estimation of individual risk is not needed for ARAIs, or any
other risk assessment method, to provide great benefit. If groups of
individuals with high and low propensities for violence recidivism can
be distinguished, and courts act upon such distinctions, recidivism
will decline to the extent that groups most prone to violence are
incapacitated, and infringements upon those least so prone are
minimized. And both society and offenders will be better served even
if we cannot be sure, based on tight statistical intervals, from
precisely which individual offenders this betterment derives.

CoHaMi's technical statistical arguments against actuarial risk
estimation are simply fallacious.  Specifically, HMC's Wilson-based
individual risk intervals, and CM1's from misapplication of a linear
regression prediction interval formula to logistic regression, direct
statistically improper computations towards an unachievable goal.
These intervals are meaningless.  HC's individual margins of error are
confidence intervals for the group risks of offenders with specific
values of the ARAI they have constructed for the purpose of their
paper.  Their widths, and those of HMC's Wilson-based and CMC's
logistic regression based group intervals, can be arbitrarily narrowed
by increasing the sizes of the samples on which they are based.  While
these widths might be used to argue that ARAIs should be developed or
at least validated using larger samples than have been used, such a
criticism would apply only to how ARAIs have been implemented, and
casts no shadow on the validity of the ARAI approach to providing
assistance to courts.

Although CoHaMi's critique based on individual risks is invalid, the
notion of individual risk itself need not be discarded. Statistical
estimation of individual risks for outcomes such as migraine headaches
or epileptic seizures is realistic.  In other circumstances, the
philosophically-minded might legitimately debate the ontological
status of a latent individual risk that can neither be estimated nor
corroborated from data. But whether \emph{ab initio} clinical
judgments, or clinically-generated modifications to individualized
actuarial base risks, can mitigate violence recidivism more
effectively and fairly than well-individualized, precisely-estimated
actuarial risks alone, is an empirical question not resolvable by
philosophical or methodological considerations or dogmatic
argumentation. We would not expect many to aver that knowledge of
physical incapacity or a brain lesion must always be disregarded in
favor of statistical risk assessment blind to such information, or
that correlates of subsequent violence deserve no consideration in the
face of a half century of literature, covering many areas of human
health and behavior, showing statistical predictions in other fields
have usually matched or outperformed expert clinical judgments
\citep{Meehl,GroveMeta,GroveMeehl}, perhaps because humans are
internally wired to impute narrative to data we encounter, and are
prone to overdo it \citep{Kahneman}.

Approaches might be compared empirically, but this requires great
care.  Anecdotal comparison is perilous.  Case histories where
clinical judgment appears superior to statistical classification are
apt to become evident \emph{ex post facto} more frequently than those
where clinical judgment has failed.  Any benefits of statistical
classification may well be distributed anonymously, although society
and individuals are no less benefited despite this anonymity.  On the
other hand, short of controlled experimentation it may be extremely
difficult to design statistical comparisons insulated from bias and
confounding, while social experimentation is expensive and may be
impractical.  A recent meta-analysis of empirical evaluations of ARAIs
found considerable heterogeneity and overall mixed results, but did
not assess or screen for study quality \citep{Fazel}.  Further
meta-analysis by these authors suggests this literature exhibits
authorship bias, and criticizes ARAI developers for not disclosing
their interests when authoring evaluation studies \citep{Singh}.

Democratic societies frequently subordinate utilitarian considerations
to social values, so prediction accuracy is not the sole legitimate
criterion for choice of risk assessment approach.  Thus, heavy
reliance on ARAIs can be questioned on other grounds.  We note several
(see \cite{Slobogin} for a general comparative review of risk
assessment approaches). Individualized risk framed as a statistical
property of a reference group may seem less appropriate than
clinically-formulated individual risks to resolving particular cases.
Statistical assessments have the potential to embed discriminatory
practice in computer code \citep{Starr}. ARAIs provide individualized
risk assessments, but such individualizations and the assessments they
produce are not unique.  Different instruments may in principle
disagree on their classifications of individuals.  Objective reasons
to choose one instrument over another in such circumstances may not be
at hand.

Specific ARAIs may be open to statistical criticisms based on methods
used in their development.  Confidence intervals for group risks
formed from the same data used to choose and weight ARAI components
will tend to be narrower than intervals formed more appropriately
using data from a new sample.  Small samples may yield less precise
intervals for individualized risks than seem advisable to inform the
consequential decisions for which they are used.

Even a validated non-discriminatory ARAI, with clear superiority to
clinical judgment in its original context, is subject to the same
questions of external validity that arise in all generalizations of
population research findings. Additional substantiation may be
warranted for application to subjects in socioeconomic environments,
communities, and cultures substantially different from where the
instrument was developed, for instance in older vs. younger or highly
rural vs. urban communities.  Risk assessments may ``age-out'' over
time and thus require ``refreshment,'' as the overall incidence of
violence rises or falls due to general influences on the culture and
specific changes affecting violence-prone strata such as young males.

The relative weights to give such varied considerations are properly
functions of social policy, not statistical inference. We conclude
that while proponents and detractors of ARAIs may have cogent
arguments to debate and for policymakers to weigh, CoHaMi's specious
statistical demonstrations are not among them.

\section{Acknowledgement}
This work was partially supported by the MacArthur Foundation Research
Network on Law and Neuroscience, and its task force on
group-to-individual (G2i) inference, charged with developing legally
practical and scientifically valid guidelines for law-relevant
inferences to individuals from group neuroscientific data, through
which the authors encountered HMC and its sequelae. G2i task force
members have provided numerous helpful comments on this manuscript,
but the authors are solely responsible for its content.

\newpage

\section{Appendix: SAS Code for Tables 2, 3, and Figure 1}
-------------------------------------------------------------------------------\\
*VRAG DATA;\\
data VRAG;\\
input Category Total Recidivists @@;\\
datalines;\\
1 11 0 2 71 6 3 101 12 4 111 19 5 116 41 6 96 42 7 74 41 8 29 22 9 9 9\\
;\\
-------------------------------------------------------------------------------\\
*STATIC-99 DATA;\\
data STATIC99;\\
input Category Total Recidivists @@;\\
datalines;\\
1 11 0 2 71 6 3 101 12 4 111 19 5 116 41 6 96 42 7 74 41 8 29 22 9 9 9\\
;\\
-------------------------------------------------------------------------------\\
LOGISTIC MODEL FOR VRAG DATA;\\
proc logistic descending data=VRAG;\\
label Category='VRAG Score Stratum';\\
*FIT MODEL;\\
model Recidivists/Total=Category;\\
*CALCULATE 95\% LOGISTIC REGRESSION INTERVALS;\\
output out=pre$\_$VRAG$\_$95 predprobs=(i) lower=Lower$\_$CL$\_$95 upper=Upper$\_$CL$\_$95/alpha=0.05;\\
*PLOT FIGURE 1a;\\
effectplot/predlabel='VRAG-Based Individualized Risk Estimate';\\
run;\\
proc logistic descending data=VRAG;\\
label Category='VRAG Score Stratum';\\
model Recidivists/Total=Category;\\
*CALCULATE 80\% LOGISTIC REGRESSION INTERVALS;\\
output out=pre$\_$VRAG$\_$80 predprobs=(i) lower=Lower$\_$CL$\_$80 upper=Upper$\_$CL$\_$80/alpha=0.2;\\
run;\\
*PRINT LOGISTIC REGRESSION INTERVALS FOR TABLE 2;\\
proc print data=pre$\_$VRAG$\_$95;\\
var Category Lower$\_$CL$\_$95 Upper$\_$CL$\_$95;\\
proc print data=pre$\_$VRAG$\_$80;\\
var Category Lower$\_$CL$\_$80 Upper$\_$CL$\_$80;\\
run;\\
-------------------------------------------------------------------------------------------------------------------------------;\\
*FOR TABLE 3 AND FIGURE 1b, CHANGE THE DATA SET AND VARIABLE NAMES.;\\
-------------------------------------------------------------------------------------------------------------------------------;\\

\newpage

\begin{table}[htdp]
  \caption{Estimates of Risk for Groups and Individuals With the Violence Risk Appraisal Guide}
  \begin{center}
    \begin{tabular}{|c|c|c|c|} \hline Category&Number of people&Number
      of Recidivists&Proportion of Recidivists \\ \hline 1&11&0&0.00
      \\ \hline 2&71&6&0.08 \\ \hline 3&101&12&0.12 \\ \hline
      4&111&19&0.17 \\ \hline 5&116&41&0.35 \\ \hline 6&96&42&0.44 \\
      \hline 7&74&41&0.55 \\ \hline 8&29&22&0.76 \\ \hline 9&9&9&1.00
      \\ \hline
    \end{tabular}
  \end{center}
  \label{default}
\end{table}%

\newpage

\begin{table}[htdp]
  \vspace{-0.5in}
  \caption{Confidence Intervals for Group Risks for Nine VRAG Score Categories, From Wilson's Method (via HMC) and Equally-Spaced Logistic Regression}
  \begin{center}
    \begin{tabular}{|c|c|c|c|} \hline
      & &\multicolumn{2}{c|}{Logistic Regression Intervals} \\
      VRAG Score Category &95\% Wilson Interval (From
      HMC)&\hspace{0.65cm} 95\%\hspace{0.65cm} & 80\% \\ \hline
      1&0.00-0.26&0.02-0.06&0.03-0.05 \\ \hline
      2&0.04-0.17&0.05-0.10&0.05-0.09 \\ \hline
      3&0.07-0.20&0.09-0.16&0.10-0.14 \\ \hline
      4&0.11-0.25&0.16-0.24&0.17-0.22\\ \hline
      5&0.27-0.44&0.27-0.35&0.28-0.34 \\ \hline
      6&0.34-0.54&0.40-0.51&0.42-0.49 \\ \hline
      7&0.44-0.66&0.53-0.67&0.56-0.65 \\ \hline
      8&0.58-0.88&0.66-0.80&0.68-0.78 \\ \hline
      9&0.70-1.00&0.76-0.89&0.79-0.88 \\ \hline
    \end{tabular}
  \end{center}
  \label{default}
\end{table}%

\newpage

\begin{table}[htdp]
  \caption{Confidence Intervals for Group Risks for Nine STATIC-99 Score Categories, From Wilson's Method (via HMC) and Equally-Spaced Logistic Regression}
  \begin{center}
    \begin{tabular}{|c|c|c|c|} \hline
      &&\multicolumn{2}{c|}{Logistic Regression Intervals} \\
      STATIC-99 Score Category &95\% Wilson Interval (From
      HMC)&\hspace{0.6cm} 95\%\hspace{0.6cm} & 80\% \\ \hline
      0&0.08-0.21&0.05-0.10&0.06-0.09 \\ \hline
      1&0.04-0.12&0.09-0.14&0.10-0.13 \\ \hline
      2&0.12-0.22&0.14-0.19&0.14-0.18 \\ \hline
      3&0.14-0.25&0.20-0.26&0.21-0.25 \\ \hline
      4&0.30-0.43&0.28-0.35&0.29-0.33 \\ \hline
      5&0.31-0.50&0.37-0.46&0..38-0.44 \\ \hline
      6+&0.43-0.60&0.46-0.58&0.48-0.56 \\ \hline
    \end{tabular}
  \end{center}
  \label{default}
\end{table}%

\newpage

\begin{table}[htdp]
  \caption{Values of 95\% Confidence Interval Bounds from Wilson's Formula (1), for Several Real and Hypothetical Sampling Scenarios With $\hat{\theta}=13\%$ Reoffending}
  \begin{center}
    \begin{tabular}{|l|c|c|c|c|c|} \hline
      \multicolumn{2}{|l|}{Offender Sample
        Scenario}&\multicolumn{2}{c|}{Reoffender?}&95\% CI
      (\%)&Ontological \\ \cline{1-4} Source&Size&Yes&No&From
      (1)&Class\\ \hline \citealt[Table 5]{STATIC-99}, via
      HMC&107&14&93&8-21&Observed\\ \hline \citealt[Table 7]{RM2000},
      via CM3&167&22&145&9-19&Observed\\ \hline
      Imrey/Dawid&100,000&13,000&87,000&12.8-13.2&Observable\\ \hline
      CM3&50&6.5&43.5&6-25&Fictitious\\ \hline
      CM3&10&1.3&8.7&3-44&Fictitious\\ \hline
      CM3&5&0.65&4.35&2-56&Fictitious\\ \hline HMC and
      CM3&1&0.13&0.87&0-84&Fictitious\\ \hline
    \end{tabular}
  \end{center}
  \label{default}
\end{table}

\newpage

\begin{figure}[h]
  \centering \subcaptionbox{VRAG Data}
  {\includegraphics[width=6in]{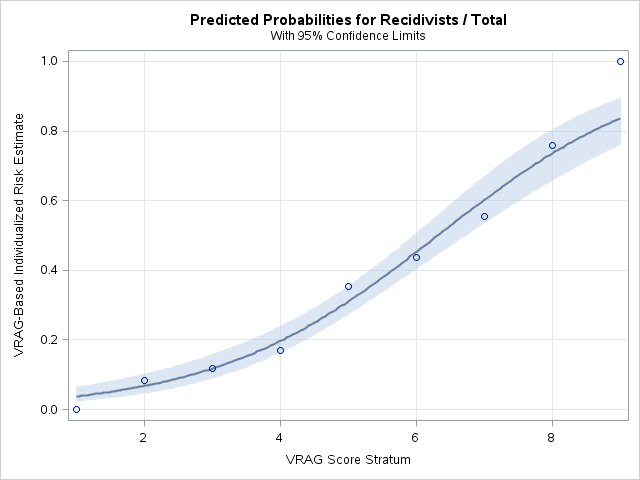}}
  \caption{Observed and fitted violence recidivism risks and 95\%
    confidence bounds from linear logistic regression with equidistant
    categories on data used by HMC.}
\end{figure}

\newpage

\begin{figure}[h]
  \setcounter{figure}{0} \centering \subcaptionbox{STATIC-99 Data}
  {\includegraphics[width=6in]{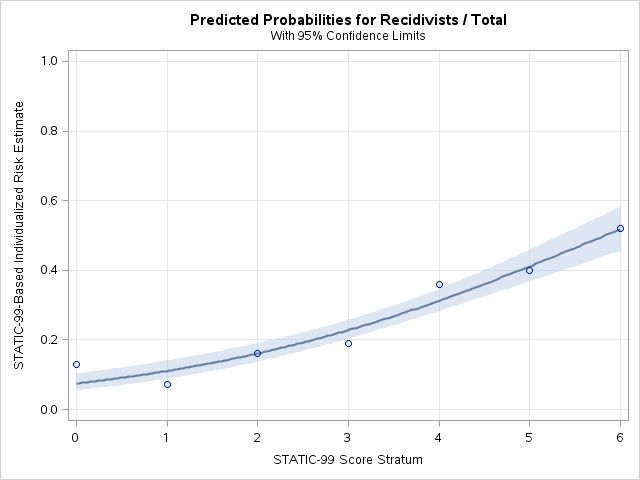}}
  \caption{Observed and fitted violence recidivism risks and 95\%
    confidence bounds from linear logistic regression with equidistant
    categories on data used by HMC.}
\end{figure}

\newpage

\bibliographystyle{apa}

\end{document}